\documentclass{article}
\usepackage{times}
\usepackage{amsfonts}
\usepackage{graphicx}
\usepackage[pdfmark]{hyperref}
\begin{document}
\noindent
{\Large AN ENTROPIC PICTURE OF EMERGENT QUANTUM MECHANICS}
\vskip1cm
\noindent
{\bf D. Acosta$^{1,a}$, P. Fern\'andez de C\'ordoba$^{2,b}$, J.M. Isidro$^{2,c}$ and J.L.G. Santander$^{3,d}$}\\
${}^{1}$Departamento de Matem\'aticas, Universidad de Pinar del R\'{\i}o,\\ Pinar del R\'{\i}o, Cuba\\
${}^{2}$Instituto Universitario de Matem\'atica Pura y Aplicada,\\ Universidad Polit\'ecnica de Valencia, Valencia 46022, Spain\\
${}^{3}$C\'atedra Energesis de Tecnolog\'{\i}a Interdisciplinar, Universidad Cat\'olica de Valencia,\\ C/ Guillem de Castro 94, Valencia 46003, Spain\\
${}^{a}${\tt dago@mat.upr.edu.cu}, ${}^{b}${\tt pfernandez@mat.upv.es}\\
${}^{c}${\tt joissan@mat.upv.es}, ${}^{d}${\tt martinez.gonzalez@ucv.es} \\
\vskip.5cm
\noindent
{\bf Abstract}  Quantum mechanics emerges {\it \`a la}\/ Verlinde from a foliation of $\mathbb{R}^3$ by holographic screens, when regarding the latter as entropy reservoirs that a particle can exchange entropy with. This entropy is quantised in units of Boltzmann's constant $k_B$. The holographic screens can be treated thermodynamically as stretched membranes. On that side of a holographic screen where spacetime has already emerged, the {\it energy representation}\/ of thermodynamics gives rise to the usual quantum mechanics. A knowledge of the different surface densities of entropy flow across all screens is equivalent to a knowledge of the quantum--mechanical wavefunction on $\mathbb{R}^3$. The {\it entropy representation}\/ of thermodynamics, as applied to a screen, can be used to describe quantum mechanics in the absence of spacetime, that is, quantum mechanics beyond a holographic screen, where spacetime has not yet emerged. Our approach can be regarded as a formal derivation of Planck's constant $\hbar$ from Boltzmann's constant $k_B$.

\tableofcontents

\section{Introduction}\label{intro}

Groundbreaking advances in our understanding of gravity have led to profound new insights into its nature (see \cite{PADDY00, PADDY0, PADDY1, PADDY2, PADDY3, VERLINDE} and refs. therein).  Perhaps the most relevant insight is the recognition that gravity  cannot a fundamental force, but rather must be an effective description of some underlying degrees of freedom.  As such, gravity is amenable to a thermodynamical description. Although this fact had already been suspected for some time \cite{HAWKINGETAL, BEKENSTEIN, HAWKING, UNRUH, JACOBSON, THOOFT0}, it is only more recently that it has been given due attention. The derivation of Newton's laws of motion and of Einstein's gravity, presented in ref. \cite{VERLINDE} from an entropic perspective, has triggered off an avalanche of research into the subject, ensueing papers being too numerous to quote here in detail; see however \cite{MUNK1, CARAVELLI, TIAN, CULETU, MODESTO, GHOSH1, GHOSH2}. A feature of these developments is that, while offering insights into the quantum structure of spacetime, the treatment is largely classical, in that no specific microscopic model of spacetime is assumed. In other words, these developments refer not to the (microscopic) statistical mechanics of gravity and spacetime, but to its (macroscopic) thermodynamics instead. In this sense, notions usually considered to be {\it a priori}\/, such as inertia, force and spacetime,  appear as phenomena arising from some underlying theory whose minuti\ae\/ are largely unknown---but fortunately also irrelevant for a thermodynamical description.  Such {\it emergent}\/ phenomena are no longer {\it a priori}\/, but derived. We refer readers to the comprehensive overview of emergent physics presented in the nice book \cite{CARROLL1}. Spacetime itself appears as an emergent phenomenon, with the holographic principle playing a key role \cite{THOOFT1, SUSSKIND}. Developments in string theory also point in this direction \cite{BLAUTHEISEN, SEIBERG}.

It has also been conjectured that quantum mechanics itself must be an emergent theory  \cite{NELSON, ADLER2, SMOLIN, THOOFT2, THOOFT3, ELZE1, ELZE2, ELZE3, ELZE4, KHRENNIKOV}; see also \cite{MATONE1, MATONE2, CARROLL0, CARROLL5, CARROLL2, CARROLL3, CARROLL4, KOCH} for its close link with gravity theories, and \cite{GROESSING3, GROESSING0, GROESSING1, GROESSING2} for an interpretation in thermodynamical terms. The guiding principle at work in many of these approaches is the notion that quantum mechanics provides some coarse--grained description of an underlying deterministic theory. In some of these models \cite{THOOFT2}, quantum states arise as equivalence classes of classical, deterministic states, the latter being grouped together into equivalence classes, or quantum states, due to our ignorance of the full microscopic description. Quantisation thus appears to be some kind of dissipation mechanism for information. In the presence of dissipation, entropy immediately comes to mind \cite{CATICHA1, CATICHA2, CATICHA3}.

Thus the two research lines mentioned above, gravity and quantum mechanics, share the common feature of being effective, thermodynamical descriptions of their respective underlying theories. It is the purpose of this paper to develop an approach to emergent quantum mechanics from the {\it entropic}\/ point of view pioneered in ref. \cite{VERLINDE}, with a quantum--mechanical particle replacing the classical particle considered in ref. \cite{VERLINDE}. Additionally, this will contribute towards clarifying the role played by Planck's constant $\hbar$ in the entropic derivation of classical  gravity (Newton's and Einstein's) presented in \cite{VERLINDE}. Indeed,
our results can be regarded as an entropic derivation of Planck's constant $\hbar$ from Boltzmann's constant $k_B$---at least conceptually if not numerically.
Altogether, our approach will provide us with {\it a holographic, entropic picture of emergent quantum mechanics}\/.

{}Finally let us say a word on notation. Awkward though the presence of $\hbar, c, G, k_B$ in our equations may seem, our purpose of exhibiting how $\hbar$ emerges from $k_B$ renders natural units inconvenient. Quantum operators will be denoted as $\hat f$, with $f$ being the corresponding classical function.

\section{Holographic screens as entropy reservoirs}\label{hsaser}

\subsection{A quantum of entropy}\label{aqoe}

The starting point in ref. \cite{VERLINDE} is a classical point particle of mass $M$ approaching a holographic screen ${\cal S}$, from that side of the latter on which spacetime has already emerged. At a distance from ${\cal S}$ equal to 1 Compton length, the particle causes the entropy $S$ of the screen to increase by the amount
\begin{equation}
\Delta S=2\pi k_B,
\label{entropia}
\end{equation}
where $k_B$ is Boltzmann's constant. The above can also be understood as meaning that $2\pi k_B$ is the {\it quantum}\/ by which the entropy of the screen increases, whenever a particle crosses ${\cal S}$. The factor $2\pi$ on the right--hand side is conventional. Relevant is only the fact that the entropy increase of the screen appears quantised in units of $k_B$.

We call {\it bright}\/ that side of the holographic screen  on which spacetime has already emerged, whereas the other side might well be termed {\it dark}\/. One can also think of the holographic screen as being the horizon of some suitably picked observer ${\cal O}$ in spacetime. For example, in the relativistic case, one can think of this observer as being a Rindler observer. The dark side might well be identified with the screen itself, as there is literally no spacetime beyond the bright side---this assertion is  to be understood as relative to the corresponding observer, since different observers might perceive different horizons. In this way, for each fixed value of the time variable,  a collection of observers ${\cal O}_{j}$, with the index $j$ running over some (continuous) set ${\cal J}$, gives rise to a foliation of 3--space by 2--dimensional holographic screens ${\cal S}_j$: $\mathbb{R}^3=\cup_{j\in {\cal J}}{\cal S}_j$. For reasons to be explained presently we will mostly restrict our attention to potentials such that the ${\cal S}_j$ are all closed surfaces; we denote the finite volume they enclose by ${\cal V}_j$, so $\partial {\cal V}_j={\cal S}_j$.

\subsection{Two thermodynamical representations}\label{ttre}

We will take (\ref{entropia}) to hold for a quantum particle as well. A quantum particle hitting the holographic screen\footnote{Due to quantum delocalisation, statements such as {\it a quantum particle hitting the holographic screen}\/ must be understood as meaning {\it a quantum--mechanical wavepacket, a substantial part of which has nonzero overlap with the screen.}} exchanges entropy with the latter, {\it i.e.}, the wavefunction $\psi$ exchanges information with ${\cal S}$. Just as information is quantised in terms of bits, so is entropy quantised, as per eqn. (\ref{entropia}). The only requirement on this exchange is that the holographic screen act as an entropy reservoir.  (See refs. \cite{RESERVA, MUNK2} for related proposals, with the mechanical action integral replacing the entropy).

Describing the quantum particle on the bright side of the screen we have the standard wavefunction $\psi_+$, depending on the spacetime coordinates and obeying the usual laws of quantum mechanics. On the other hand, the {\it entropic}\/ wavefunction $\psi_-$ describes the same quantum particle, as seen by an observer on the dark side of the holographic screen. If imagining an observer on the dark side of ${\cal S}$, where spacetime has not yet emerged, raises some concern, one can also think of $\psi_-$ as being related, in a way to be made precise below, to the flow of entropy across the horizon ${\cal S}$, as measured by an observer on the bright side of the same horizon.

Our goal is to describe the laws of {\it entropic quantum mechanics}\/, that is, the laws satisfied by the entropic wavefunction $\psi_-$, and to place them in correspondence with those satisfied by the standard wavefunction $\psi_+$ on spacetime. The relevant thermodynamical formalism needed here can be found, {\it e.g.}, in the classic textbook \cite{CALLAN}. However, for later use, let us briefly summarise a few basics. Any given thermodynamical system can be completely described if one knows its {\it fundamental equation}\/. The latter contains all the thermodynamical information one can obtain about the system. The fundamental equation can be expressed in either of two equivalent ways, respectively called the {\it energy representation}\/ and the {\it entropy representation}\/. In the energy representation one has a fundamental equation $E=E(S, \ldots)$, where the energy $E$ is a function of the entropy $S$, plus of whatever additional variables may be required. In the entropy representation one solves for the entropy in terms of the energy to obtain a fundamental equation $S=S(E, \ldots)$.

As an example, let there be just one extensive parameter, the volume $V$. Then the fundamental equation in the entropy representation will be an expression of the form $S=S(E, V)$, hence ${\rm d}S=\left(\partial S/\partial E\right){\rm d}E+\left(\partial S/\partial V\right){\rm d}V$. We know that $\delta Q=T{\rm d}S$, while the first law of thermodynamics reads, in this case,  $\delta Q={\rm d}E+p{\rm d}V$, with $p$ the pressure. It follows that $T^{-1}=\partial S/\partial E$ and $p=T\left(\partial S/\partial V\right)$. This latter equation is the equation of state. For example, in the case of an ideal gas we have $S(E,V)=k_B\ln \left(V/V_0\right)+f(E)$, with $f(E)$ a certain function of the energy and $V_0$ a reference volume (that can be regarded as a constant contribution to $S$ and thus neglected). It follows from $\partial S/\partial V=k_B V^{-1}$ that $pV$ is proportional to $T$, as expected of an ideal gas.

In a sense to be made more precise presently, the bright side of the holographic screen corresponds to the energy representation, while the dark side corresponds to the entropy representation. Thus the energy representation will give us quantum mechanics on spacetime as we know it. One must bear in mind, however, that standard thermodynamical systems admit both representations (energy and entropy) simultaneously, which representation one uses being just a matter of choice. In our case this choice is dictated, for each fixed observer, by that side of the screen on which the observer wants to study quantum mechanics. For example there is no energy variable on the dark side, as there is no time variable, but an observer can assign the screen an entropy, measuring the observer's ignorance of what happens beyond the screen. By the same token, on the bright side we have an energy but there is no entropy\footnote{We are considering the simplified case of a pure quantum state. Were our quantum state to be described by a density matrix, there would of course be an entropy associated.}. In this case these two representations cannot be simultaneous.

The situation just described changes somewhat as soon as one considers two or more observers, each one of them perceiving a different horizon or holographic screen. Consider, for simplicity, two observers ${\cal O}_1, {\cal O}_2$ with their respective screens ${\cal S}_1, {\cal S}_2$, and assume the latter to be such that ${\cal S}_2$ gets beyond ${\cal S}_1$, in the sense that ${\cal S}_2$ encloses more emerged volume than ${\cal S}_1$. That is, the portion of emerged spacetime perceived by ${\cal O}_2$ includes all that perceived by ${\cal O}_1$, plus some volume that remains on the dark side of ${\cal S}_1$.  Call ${\cal V}_{12}$ this portion of spacetime that appears dark to ${\cal O}_1$ but bright to ${\cal O}_2$. Clearly, quantum mechanics on ${\cal V}_{12}$ will be described in the energy representation by ${\cal O}_2$ and in the entropy representation by ${\cal O}_1$. In this case the two representations can coexist simultaneously---not as corresponding to one observer, as in standard thermodynamics, but each one of them as {\it pertaining to a different observer}\/.

The differences just mentioned, as well as some more that will arise along the way, set us somewhat apart from the standard thermodynamical formalism. Nevertheless, the thermodynamical analogy can be quite useful if one bears these differences in mind.

\subsection{A holographic dictionary}\label{aholdic}

Let us recall that one can formulate a {\it holographic dictionary}\/ between gravitation, on the one hand, and thermodynamics, on the other \cite{PADDY00, PADDY0, PADDY1, PADDY2, PADDY3}. Let $V_G$ denote the gravitational potential created by a total mass $M=\int_{\cal V}{\rm d}^3V\,\rho_M$ within the volume ${\cal V}$ enclosed by the holographic screen ${\cal S}=\partial{\cal V}$. Then the following two statements are equivalent \cite{VERLINDE, SABINE}:\\
{\it i)} there exists a gravitational potential $V_G$ satisfying Poisson's equation $\nabla^2 V_G=4\pi G\rho_M$, such that a test mass $m$ in the background field created by the mass distribution $\rho_M$ experiences a force ${\bf F}=-m\nabla V_G$;\\
{\it ii)} given a foliation of 3--space by holographic screens, $\mathbb{R}^3=\cup_{j\in {\cal J}}{\cal S}_j$, there are two scalar quantities, called entropy $S$ and temperature $T$, such that the force acting on a test mass $m$ is given by $F\delta x=\int_{\cal S}T\delta {\rm d}S$. The latter integral is taken over a screen that does not enclose $m$.\\
Moreover, the thermodynamical equivalent of the gravitational theory includes the following {\it dictionary entries}\/ \cite{VERLINDE}:
\begin{equation}
\frac{1}{k_B}S(x)=\frac{-1}{4\hbar c L_P^2}V_G(x)A(V_G(x)),
\label{sabinebiene}
\end{equation}
\begin{equation}
2\pi k_BT(x)=\frac{{\rm d}V_G}{{\rm d}n},
\label{tivi}
\end{equation}
\begin{equation}
\frac{k_B}{2}\int_{\cal S}{\rm d}^2a\, T=L_P^2Mc^2.
\label{defitermo}
\end{equation}
In (\ref{sabinebiene}), (\ref{tivi}) and (\ref{defitermo}) we have placed all thermodynamical quantities on the left, while their mechanical analogues are on the right. As in ref. \cite{VERLINDE}, the area element ${\rm d}^2a$ on ${\cal S}$ is related to the infinitesimal number of bits ${\rm d}N$ on it through ${\rm d}^2a=L_P^{2}{\rm d}N$. We denote the area of the equipotential surface passing through the point $x$ by $A(V_G(x))$, while ${\rm d}V_G/{\rm d} n$ denotes the derivative of $V_G$ along the normal direction to the same equipotential. The above expressions tell us how, given a gravitational potential $V_G(x)$ and its normal derivative ${\rm d}V_G/{\rm d} n$, the entropy $S$ and the temperature $T$ can be defined {\it as functions of space}\/.

Specifically, eqn. (\ref{sabinebiene}) expresses the proportionality between the area $A$ of the screen ${\cal S}$ and the entropy $S$ it contains. This porportionality implies that gravitational equipotential surfaces get translated, by the holographic dictionary, as {\it isoentropic surfaces}\/, above called holographic screens ${\cal S}$.

Equation (\ref{tivi}) expresses the Unruh effect: an accelerated observer experiences the vacuum of an inertial observer as a thermal bath at a temperature $T$ that is proportional to the observer's acceleration ${\rm d}V_G/{\rm d}n$.

{}Finally, eqn. (\ref{defitermo}) expresses the first law of thermodynamics and the equipartition theorem. The right--hand side of (\ref{defitermo}) equals the total rest energy of the mass enclosed by the volume ${\cal V}$, while the left--hand side expresses the same energy content as spread over the bits of the screen ${\cal S}=\partial {\cal V}$, each one of them carrying an energy $k_BT/2$. It is worthwhile noting that equipartition need not be postulated. Starting from (\ref{tivi}) one can in fact prove the following form of the equipartition theorem:
\begin{equation}
\frac{k_B}{2}\int_{\cal S}{\rm d}^2a\, T=\frac{A({\cal S})}{4\pi}U({\cal S}), \qquad A({\cal S})=\int_{\cal S}{\rm d}^2a.
\label{defitermonova}
\end{equation}
The details leading up to (\ref{defitermonova}) from (\ref{tivi}) will be given in section \ref{tfeteosaeq}. Above, $U$ can be an arbitrary potential energy\footnote{The gravitational potential $V_G$ appearing above is the gravitational energy $U_G$ per unit test mass $m$.}. We will henceforth mean eqn. (\ref{defitermonova}) when referring to the first law and the equipartition theorem. In all the above we are treating the area as a continuous variable, but in fact it is quantised \cite{VERLINDE}. If $N({\cal S})$ denotes the number of bits of the screen ${\cal S}$, then
\begin{equation}
A({\cal S})=N({\cal S})L_P^2.
\label{numm}
\end{equation}
However, in the limit $N\to\infty$, when $\Delta N/N<<1$, this approximation of the area by a continuous variable is accurate enough. We will see later on that letting $N\to\infty$ is equivalent to the semiclassical limit in quantum mechanics.

We intend to write a holographic dictionary between quantum mechanics, on the one hand, and thermodynamics, on the other. This implies that we will need to generalise eqns.  (\ref{sabinebiene}), (\ref{tivi}) and (\ref{defitermonova}) so as to adapt them to our quantum--mechanical setup.  Thus we will replace the classical particle of \cite{VERLINDE} with a quantum particle, subject to some potential energy $U$ of nongravitational origin.

\section{The energy representation}\label{tenrep}

Let $H=K+U$ be the classical Hamiltonian function on $\mathbb{R}^3$ whose quantisation leads to the quantum Hamiltonian operator $\hat H=\hat K+\hat U$ that governs our quantum particle. The Hamiltonian $\hat H$ will be assumed to possess normalisable states. This condition on the potential was already reflected in the gravitational case of eqn. (\ref{sabinebiene}), where the negative sign of the gravitational potential led to a positive definite entropy.

On the bright side of the screen, spacetime has already emerged. This gives us the energy representation of quantum mechanics---the one we are used to: a time variable with a conserved Noether charge, the energy, and wavefunctions depending on the spacetime coordinates. We have the uncertainty relation
\begin{equation}
\Delta \hat Q\,\Delta \hat P\geq \frac{\hbar}{2}.
\label{ineq}
\end{equation}
In the semiclassical limit we have a wavefunction
\begin{equation}
\psi_+=\exp\left(\frac{{\rm i}}{\hbar}I\right),
\label{sabemos}
\end{equation}
where $I=\int{\rm d}t L$ is the action integral satisfying the Hamilton--Jacobi equation.

Let ${\cal V}$ denote the finite portion of 3--space bounded by the closed holographic screen ${\cal S}=\partial{\cal V}$. We can now posit the quantum--mechanical analogues of eqns.  (\ref{sabinebiene}), (\ref{tivi}) and (\ref{defitermonova}). In the energy representation these analogues read, respectively,
\begin{equation}
\frac{1}{k_B}\hat S(x)=\frac{1}{4\hbar c L_P}A(U(x)){\vert\hat U(x)\vert},
\label{lieber}
\end{equation}
\begin{equation}
2\pi k_B\hat T(x)=L_P\frac{{\rm d}\hat U}{{\rm d}n},
\label{tivix}
\end{equation}
\begin{equation}
\frac{k_B}{2}\int_{\cal S}{\rm d}^2a\, \hat T=\frac{A({\cal S})}{4\pi}\hat U ({\cal S}).
\label{deficuanto}
\end{equation}
Some comments are in order. We are considering the nonrelativistic limit, in which the rest energy of the particle can be ignored. We also neglect all gravitational effects, relativistic or not; we will limit ourselves to the external potential $\hat U$. Quantum operators such as $\hat U$, initially defined to act on wavefunctions in $L^2(\mathbb{R}^3)$, must now be restricted to act on wavefunctions in $L^2({\cal V})$. Denote this restriction by $\hat U_{\cal V}$. By definition, its matrix elements $\langle f_+\vert\hat U_{\cal V}\vert g_+\rangle$ are
\begin{equation}
\langle f_+\vert\hat U_{\cal V}\vert g_+\rangle:=\int_{\cal V}{\rm d}^3V f_+^*\hat U g_+,
\label{porcion}
\end{equation}
the integral extending over the finite volume ${\cal V}$ instead of all $\mathbb{R}^3$. For simplicity we have suppressed the subindex ${}_{\cal V}$ in  (\ref{lieber}), (\ref{tivix}) and (\ref{deficuanto}), but it must be understood that all operators are to be restricted as specified.

The right--hand side of (\ref{lieber}) deserves more attention. $\vert \hat U\vert$ denotes the operator whose matrix elements are the absolute values of those of $\hat U$. Taking the absolute value ensures that the entropy is positive definite, given that the potential $U$ need not have a constant sign, contrary to the gravitational case of (\ref{sabinebiene}).

It will also be observed that no carets stand above $A(U(x))$, $A({\cal S})$, because they are c--numbers. They denote the area of the equipotential surface passing through the point $x$ and the are of the screen ${\cal S}$, respectively. Also, the integral on the left--hand side of (\ref{deficuanto}) is a standard surface integral, even if the integrand is the operator $\hat T$, because the latter depends on the c--number--valued coordinate functions $x$.

As a final remark, let us point out that the above equations (\ref{lieber}), (\ref{tivix}) and (\ref{deficuanto}), as well as their classical counterparts (\ref{sabinebiene}), (\ref{tivi}) and (\ref{defitermonova}), are correctly understood as being expressed in the energy representation of thermodynamics. This is so despite the fact that one writes the entropy as an explicit function of the potential energy---would this not be the defining property of the entropy representation? The answer is negative for two reasons. First, one would need to express the entropy as a function of the total energy $H$, rather than as a function of just the potential energy $U$. Second, all the above expressions are functions defined on the emerged portion of space, where there exists a conserved Noether charge, the energy $H$, and its conjugate variable, the time $t$. The entropy representation will be introduced later on, when the absence of spacetime will make it necessary to eliminate the space dependence of quantities such as entropy and temperature. Such will be the case beyond the holographic screen.

\section{The entropy representation}\label{thentrep}

The entropy representation can also be thought of as quantum mechanics in the absence of spacetime, as we will come to recognise presently.

\subsection{Action {\it vs}\/. entropy}\label{acvsentr}

It is well known, in the theory of thermodynamical fluctuations \cite{CALLAN}, that the probability density function $d$ required to compute expectation values of thermodynamical quantities is given by the exponential of the entropy:
\begin{equation}
d=\exp\left(\frac{S}{k_B}\right).
\label{scheiss}
\end{equation}
Its square root, that one may call the amplitude for the probability density $d$, can therefore be identified with an entropic wavefunction $\psi_-^{(d)}$:
\begin{equation}
\psi_-^{(d)}=\exp\left(\frac{S}{2k_B}\right).
\label{gemuetlich}
\end{equation}
This identification is made up to a (possibly point--dependent) phase ${\rm e}^{{\rm i}\alpha}$, plus a normalisation. Comparing (\ref{gemuetlich}) with (\ref{sabemos}) we arrive at the correspondence
\begin{equation}
\frac{{\rm i}I}{\hbar}\leftrightarrow\frac{S}{2k_B}
\label{korres}
\end{equation}
between the energy representation and the entropy representation, {\it both of them taken in the semiclassical limit}\/. This amounts to the statement that quantum--mechanical fluctuations can be understood thermodynamically, at least in the semiclassical limit.

We should note that the correspondence (\ref{korres}) is holographic in nature, because the action integral $I$ is defined on space, while the entropy $S$ is defined on the screen bounding it. Moreover, the above correspondence also implies that, in the entropic representation, the semiclassical limit (the one considered in (\ref{sabemos})) corresponds to letting $k_B\to 0$.

The wavefunction (\ref{gemuetlich}) describes an {\it incoming}\/ wave, from the point of view of the screen. An {\it outgoing}\/ wave, from the point of view of the screen, would be described by $\exp\left(-S/2k_B\right)$.

It is reassuring to observe that the same correspondence (\ref{korres}) has been found in the context of gravity and black--hole thermodynamics \cite{BANERJEE1, BANERJEE2}.

\subsection{Quantum states {\it vs}\/. holographic screens}\label{qsvshs}

The equation $U(x^1, x^2, x^3)=U_0$, where $U_0$ is a constant, defines an equipotential surface in $\mathbb{R}^3$. As $U_0$ runs over all its possible values, we obtain a foliation of  $\mathbb{R}^3$ by equipotential surfaces. Following \cite{VERLINDE}, we will identify equipotential surfaces with holographic screens. Hence forces will arise as entropy gradients.

Assume that $\psi_+$ is nonvanishing at a certain point in space. Consider an infinitesimal cylinder around this point, with height $L_P$ and base area equal to the area element ${\rm d}^2a$. Motivated by the proportionality between area and entropy, already mentioned, we postulate that there is an infinitesimal entropy flow ${\rm d}S$ {\it from the particle to the area element}\/ ${\rm d}^2a$:
\begin{equation}
{\rm d}S=C \,2\pi k_BL_P \vert\psi_+\vert^2{\rm d}^2a.
\label{helement}
\end{equation}
Here $C$ is a dimensionless numerical constant, to be determined presently. A closed surface $\Sigma$ receives an entropy flux $S(\Sigma)$:
\begin{equation}
S(\Sigma)=C(\Sigma)2\pi k_BL_P \int_{{\Sigma}}{\rm d}^2a\,\vert\psi_+\vert^2.
\label{hassociat}
\end{equation}
The constant $C(\Sigma)$ will in general depend on the particular surface chosen; the latter may, but need not, be a holographic screen.  The key notion here is that the integral of the scalar field $\vert\psi_+\vert^2$ over any surface carries an entropy flow associated. When the surface $\Sigma$ actually coincides with a holographic screen ${\cal S}$, and when the latter is not a nodal surface of $\psi_+$, the constant $C({\cal S})$ may be determined by the requirement that the entropy flux from the particle to the screen equal the quantum of entropy (\ref{entropia}). Thus
\begin{equation}
\frac{1}{C({\cal S})}=L_P \int_{{\cal S}}{\rm d}^2a\,\vert\psi_+\vert^2.
\label{cequattro}
\end{equation}
We should point out the following. Given a wavefunction $\psi_+$, the probability density $\vert\psi\vert^2$ on $\mathbb{R}^3$ gives rise to a natural definition of entropy, namely, 
\begin{equation}
-k_B\int{\rm d}^3V\,\vert\psi_+\vert^2\log \vert\psi_+\vert^2.
\label{ovvio}
\end{equation}
However, (\ref{ovvio}) is the entropy associated with our uncertainty in the position of the particle in 3--space. As such it should not be confused with the entropy (\ref{hassociat}) associated with the particle traversing the surface $\Sigma$. It is this latter entropy that we are interested in.

Let us now read eqn. (\ref{cequattro}) in reverse, under the assumption that one knows the proportionality constants $C({\cal S}_j)$ for a given foliation $\mathbb{R}^3=\cup_{j\in {\cal J}}{\cal S}_j$. This amounts to a knowledge of  the integrands, {\it i.e.}, of the probability density $\vert\psi_+\vert^2$ within the surface integral (\ref{cequattro}) on each and every ${\cal S}_j$. From these tomographic sections of all probability densities {\it there emerges the complete wavefunction $\psi_+$ on all of}\/ $\mathbb{R}^3$, at least up to a (possibly point--dependent) phase ${\rm e}^{{\rm i}\alpha}$.

Thus the integrand of (\ref{cequattro}) gives the surface density of entropy flow into the holographic screen ${\cal S}_j$, and the wavefunction $\psi_+$ becomes (proportional to) the square root of this flow. The collection of all these tomographic sections of $\psi_+$ along all possible screens amounts to a knowledge of the complete wavefunction. Hence {\it a knowledge of the different surface densities of entropy flux across all possible screens is equivalent to a knowledge of the quantum--mechanical wavefunction} $\psi_+$. This is how the quantum--mechanical wavefunction $\psi_+$ emerges from the holographic screens. Close ideas concerning the wavefunction 
in relation to foliations of space have been put forward in ref. \cite{CARROLL0}.

\subsection{The entropic uncertainty principle}\label{tenuncprp}

Let us define the dimensionless variable
\begin{equation}
s:=\frac{S}{2\pi k_B},
\label{sindim}
\end{equation}
that we will call the {\it reduced entropy}\/. It is nonnegative: $s\geq 0$. For example, the semiclassical entropic wavefunction (\ref{gemuetlich}) can be expressed in terms of $s$ as $\psi_-^{(d)}(s)={\rm e}^{\pi s}$.  We can consider arbitrary functions $f(s)$ on which we let the following operators $\hat Q_S, \hat P_S$ act:
\begin{equation}
\hat  Q_Sf(s):=sf(s),\qquad \hat P_Sf(s):=2\pi k_B\frac{{\rm d}f(s)}{{\rm d}s}.
\label{koordinat}
\end{equation}
{}For reasons that will become clear presently, $\hat Q_S$ will also be called the {\it normal, or entropic, position operator}\/, while $\hat P_S$ will be called the {\it normal, or entropic, momentum}\footnote{The missing factor of $i$ in the definition of $\hat P_S$ is due to the correspondence (\ref{korres}).}\/. One finds that $ i\hat P_S$ and $\hat Q_S $ are Hermitian on $L^2\left[0,\infty\right)$. Unlike the usual case on $L^2(\mathbb{R})$, the Hermitian property of position and momentum on the semiaxis involves some nontrivial mathematical subtleties that will not be touched upon here; see \cite{THIRRING}. Now the above operators satisfy the Heisenberg algebra
\begin{equation}
[\hat Q_S, \hat P_S]=2\pi k_B{\bf 1}.
\label{heialg}
\end{equation}
Therefore the following {\it entropic uncertainty principle}\/ holds:
\begin{equation}
\Delta \hat Q_S\,\Delta \hat P_S\geq \pi k_B.
\label{ungewiss}
\end{equation}
The above uncertainty principle has been derived rather than postulated; this is in the spirit of refs. \cite{GOSSON1, GOSSON2}.

\subsection{The entropic Schroedinger equation}\label{opeeigs}

Since the screens ${\cal S}_j$ are isoentropic surfaces,  the reduced entropy $s$ can be regarded as a dimensionless coordinate orthogonal to all the ${\cal S}_j$. Multiplication by $L_P$ gives a dimensionful coordinate $\rho$:
\begin{equation}
\rho:=L_Ps.
\label{hijosdeputa}
\end{equation}
Modulo multiplication by a dimensionless numerical factor, and the possible addition of a constant, the above is an equivalent reexpression of the equation \cite{VERLINDE}
\begin{equation}
\Delta S=2\pi k_B\frac{Mc}{\hbar}\Delta x,
\label{verlindeq}
\end{equation}
where $x$ is the distance measured normally to the screen---in turn, (\ref{verlindeq}) is the same as (\ref{entropia}). We can exploit this fact if we assume that the time--independent Schroedinger equation
\begin{equation}
-\frac{\hbar^2}{2M}\nabla^2 \psi_+ + U\psi_+ = E \psi_+
\label{tise}
\end{equation}
is separable in a coordinate system that includes $\rho$ as one of its coordinate functions. So let us supplement $\rho$ with two additional coordinates $\xi, \chi$ such that the triple $\rho, \xi, \chi$ provides an orthogonal set of curvilinear coordinates\footnote{In general, $\rho, \xi, \chi$ are only local coordinates, and need not cover all of $\mathbb{R}^3$. In particular, $\xi, \chi$ need not cover a complete screen ${\cal S}_j$, nor need they be simultaneously defined on different screens ${\cal S}_j$, ${\cal S}_k$. However, to simplify our notation, we omit all the indices that would be necessary in order to take all these possibilities into account.} in which (\ref{tise}) separates as per (\ref{sepracion}) below. Then the Euclidean line element on $\mathbb{R}^3$ will be given by
\begin{equation}
{\rm d}s^2=h_{\rho}^2{\rm d}\rho^2 + h_{\xi}^2{\rm d}\xi^2 + h_{\chi}^2{\rm d}\chi^2,
\label{koorp}
\end{equation}
where the metric coefficients $h_{\rho}, h_{\xi}, h_{\chi}$ are functions of all three coordinates $\rho, \xi, \chi$. We will call $\rho$ the {\it normal coordinate}\/ to the foliation, while $\xi,\chi$ will be called {\it tangential coordinates}\/ to the foliation. A more physical terminology, based on (\ref{hijosdeputa}) and (\ref{tivix}), could be {\it entropic coordinate}\/ for $\rho$ and {\it isothermal coordinates}\/ for $\xi, \chi$.

We recall that $U$ depends only on the normal coordinate $\rho$, so equipotential surfaces are defined by  $U(\rho)=U_0$, for any constant $U_0$. The tangential dimensions $\xi, \chi$ are purely spatial constructs: they encode the geometry of the equipotential surfaces. For example, in the particular case of a Coulomb potential, or also of an isotropic harmonic oscillator, the ${\cal S}_j$ are a family of concentric spheres of increasing radii. Then $\rho$ can be identified with the usual radial coordinate $r$ on $\mathbb{R}^3$, while $\xi, \chi$ can be taken as the usual polar angles $\theta, \varphi$. In the general case $\rho, \xi, \chi$ need not coincide with any of the standard coordinate functions on $\mathbb{R}^3$. However, each screen ${\cal S}_j$ can be univocally identified by the equation $\rho=\rho_j$. The uncertainty principle (\ref{ungewiss}) holds on the phase space corresponding to $\rho$, and the operator $\hat Q_S$ defined in (\ref{koordinat}) is nothing but {\it the position operator along the normal, or entropic, coordinate}\/.

Thus separating variables as per
\begin{equation}
\psi_+(\rho,\xi,\chi)=R(\rho)Y(\xi,\chi),
\label{sepracion}
\end{equation}
and substituting into (\ref{tise}) leads to
$$
\frac{1}{h_{\rho}h_{\xi}h_{\chi}}\left[
\frac{1}{R}\frac{\partial}{\partial\rho}\left(\frac{h_{\xi}h_{\chi}}{h_{\rho}}\frac{\partial R}{\partial\rho}\right)
+\frac{1}{Y}\frac{\partial }{\partial\xi}\left(\frac{h_{\rho}h_{\chi}}{h_{\xi}}\frac{\partial Y}{\partial\xi}\right)
+\frac{1}{Y}\frac{\partial }{\partial\chi}\left(\frac{h_{\rho}h_{\xi}}{h_{\chi}}\frac{\partial Y}{\partial\chi}\right)
\right]
$$
\begin{equation}
+\frac{2M}{\hbar^2}(E-U)=0.
\label{merde}
\end{equation}
The precise way in which (\ref{merde}) separates into a $\rho$--dependent piece and a $\xi, \chi$--dependent piece cannot be written down in all generality, as it varies according to the particular choice made for $\rho, \xi, \chi$. This is due to our ignorance of the specific way in which the metric coefficients $h_{\rho}$, $h_{\xi}$, $h_{\chi}$ depend on all three variables $\rho, \xi, \chi$. One can, however, outline some general features of the final outcome. Terms involving the Laplacian $\nabla^2$ will decompose as a sum $\nabla^2_{\rho}+\nabla^2_{\xi, \chi}$, where subindices indicate the variables being differentiated in the corresponding operators. Calling the separation constant $\lambda$, there will be two separate equations. The first equation will involve the normal Laplacian $\nabla^2_{\rho}$, the potential energy $U(\rho)$, the energy eigenvalue $E$, the mass $M$ and the separation constant $\lambda$. All these elements (with the exception of $\nabla^2_{\rho}$) appear as a certain function $F$ of $\rho$:
\begin{equation}
\nabla^2_{\rho}R(\rho)+F(\rho,U(\rho),E,M,\lambda)R(\rho)=0.
\label{raddi}
\end{equation}
The unknown function $F$ is explicitly computable once a specific choice has been made for the coordinates $\xi, \chi$. The second equation involves only the tangential Laplacian $\nabla^2_{\xi, \chi}$ and the separation constant $\lambda$:
\begin{equation}
\nabla^2_{\xi, \chi}Y(\xi, \chi)+\lambda Y(\xi, \chi)=0.
\label{armonico}
\end{equation}
It is important to note that (\ref{armonico}) can be solved independently of (\ref{raddi})\footnote{Needless to say, in the case of a Coulomb field, (\ref{raddi}) becomes the standard radial wave equation, while (\ref{armonico}) becomes that satisfied by the usual spherical harmonics, with $\lambda=l(l+1)$.}. The eigenfunctions $Y(\xi,\chi)$ constitute a complete orthonormal system of eigenfunctions of the tangential Laplacian within the {\it tangential Hilbert space}\/ $L^2({\cal S}_j)$. Moreover, since we have assumed the screens to be closed surfaces, the eigenvalues $\lambda$ will be quantised. Once these eigenvalues have been determined, substitution into (\ref{raddi}) allows the latter to be completely solved.

We are finally in a position to define the entropic wavefunction $\psi_-$ in terms of its partner $\psi_+$. We take the entropic wavefunction to be the $\rho$--dependent piece in the factorisation (\ref{sepracion}),
\begin{equation}
\psi_-(\rho):=R(\rho).
\label{difamo}
\end{equation}
Clearly the {\it entropic, or normal, Hilbert space}\/ corresponding to the screen ${\cal S}_j$ will be $L^2[0, \rho_j)$. The latter is considered with respect to an integration measure that includes a certain Jacobian factor $J(\rho)$. In order to compute this Jacobian we proceed as follows. Apply the factorisation (\ref{sepracion}) to the normalisation condition for $\psi_+$ on ${\cal V}_j$:
\begin{equation}
\int_{{\cal V}_j}{\rm d}^3V\,\vert\psi_+\vert^2=\int_0^{\rho_j}{\rm d}\rho  \int_{{\cal S}_j}{\rm d}{\xi}{\rm d}\chi\,h_{\rho}h_{\xi}h_{\chi}\vert R(\rho)\vert^2\vert Y(\xi, \chi)\vert^2.
\label{mrmm}
\end{equation}
In general, the product $h_{\rho}h_{\xi}h_{\chi}$ depends on all three coordinates $\rho, \xi, \chi$. The sought--for Jacobian $J(\rho)$ equals the $\rho$--dependent factor in the integration measure after the integral over $\xi, \chi$  has been carried out.  As $\rho_j$ becomes larger and larger, we obtain the entropic Hilbert space $L^2[0, \infty)$. The latter would correspond to an observer who perceives no horizon at all, thus extending his normalisation integral (\ref{mrmm}) over all of $\mathbb{R}^3$. We will come back to the issue of the different realisations of the entropic Hilbert space ($L^2[0, \rho_j)$ {\it vs.} $L^2[0, \infty)$) in section \ref{qmitabsospct}.

In the passage form the energy representation to the entropy representation we appear to have lost the information corresponding to the holographic screens one integrates over. However the screens carry no dynamics, because the force at point $x$ is orthogonal to the screen passing through $x$. Thus a knowledge of the entropic wavefunction $\psi_-$, {\it plus of the foliation itself}\/, is equivalent to a knowledge of the wavefunction $\psi_+$ in the energy representation. That the foliation is a piece of information belonging to the entropy representation, was stated in assertion {\it ii)} of our section \ref{aholdic} following \cite{VERLINDE, SABINE}.

It remains to identify the wave equation satisfied by the entropic wavefunction $\psi_-$. Obviously this equation is (\ref{raddi}), which may thus be regarded as {\it the entropy--representation analogue of the time--independent Schroedinger equation}\/ $\hat H\psi_+=E\psi_+$ on space. Recalling (\ref{lieber}) and (\ref{hijosdeputa}), this entropic Schroedinger equation reads
\begin{equation}
\nabla^2_{s}\psi_-(s)+G(s, A(s),E,M,\lambda)\psi_-(s)=0.
\label{raddius}
\end{equation}
We have called $G(s,A(s),E,M,\lambda)$ the function that results from expressing the potential $U$ as a function of the entropy $S$ and the area $A$, and writing everything in terms of the reduced entropy $s$. As was the case with $F$ in (\ref{raddi}), the unknown function $G$ is explicitly computable once a specific choice has been made for the coordinates $\xi, \chi$.

\subsection{The fundamental equation, the equation of state, and equipartition}\label{tfeteosaeq}

In this section we will rewrite  the dictionary entries (\ref{lieber}), (\ref{tivix}) and (\ref{deficuanto}), found to hold in the energy representation, in the entropy representation. For this purpose we first need to solve the eigenvalue equation $\hat S\phi_-=S\phi_-$ on the screen, so the latter will be kept fixed. That is, we will not consider a variable surface ${\cal S}_j$ of the foliation, but  rather a specific surface corresponding to a fixed value of the index $j$. Observe also a difference in notation: $\phi$ instead of $\psi$. This is to stress the fact that, by (\ref{lieber}), entropy eigenstates $\phi$ cannot be eigenstates of the complete Hamiltonian $\hat H$, but only of the potential energy $\hat U$. Once $\hat U$ is diagonalised by a set of $\phi_+$ defined on the bright side, {\it i.e.}, once we have solved the eigenvalue equation\footnote{Obviously the $\phi_+$ are the well--known eigenfunctions of the position operator on the bright side, but this property is immaterial for our purposes.}
\begin{equation}
\hat U\phi_+=U\phi_+,
\label{usombrero}
\end{equation}
then the corresponding $\phi_-$ on the screen are defined per continuity: $\phi_-({\cal S})=\phi_+({\cal S})$. By (\ref{lieber}), the same $\phi_-$ then diagonalise $\hat S$:
\begin{equation}
\hat S\phi_-=S\phi_-, \qquad S=\frac{k_B}{4\hbar c L_P}A({\cal S}){\vert U({\cal S})\vert}.
\label{eigenentropia}
\end{equation}
Thermodynamical quantities will now arise as expectation values of operators in the entropic eigenstates $\phi_-({\cal S})$.  

We first deal with (\ref{lieber}). Clearly its reexpression in the entropy representation will be the thermodynamical fundamental equation $S=S(A)$ in the sense of ref. \cite{CALLAN}, since the extensive parameter corresponding to the holographic screen is the area $A$. Then we have
\begin{equation}
\langle \hat S\rangle=\frac{k_B}{4\hbar c L_P}A({\cal S}){\vert U({\cal S})\vert}.
\label{biacca}
\end{equation}
Availing ourselves of the freedom to pick the origin of potentials at will, let us set $\vert U({\cal S})\vert=\hbar c/L_P$. Thus
\begin{equation}
\langle \hat S\rangle=\frac{k_B}{4L_P^2}A,
\label{bhache}
\end{equation}
which is the celebrated Bekenstein--Hawking law. It arises as a thermodynamical fundamental equation in the entropy representation.

Our holographic screen is treated thermodynamically as a stretched membrane, so the generalised force conjugate to the extensive parameter $A$ is the surface tension $\sigma$. Then the equation of state corresponding to (\ref{bhache}) is
\begin{equation}
\sigma=\frac{k_B\langle \hat T\rangle}{4L_P^2}.
\label{estado}
\end{equation}
Rewrite the above as $2\pi k_B\langle \hat T\rangle=8\pi L_P^2\sigma$ and recall that $\sigma$ is the normal component of force per unit length on the screen. Since force is proportional to acceleration, the above equation of state turns out to be equivalent to the Unruh law.

{}Finally we turn to the first law of thermodynamics and the equipartition theorem. As already mentioned in section \ref{aholdic}, it turns out that the equipartition theorem can be derived from the Unruh law. Since this fact is valid both in the classical case (\ref{defitermonova}) and in its quantum counterpart (\ref{deficuanto}), the derivation being exactly the same whatever the case, we will provide the details pertaining to the derivation of (\ref{deficuanto}) from (\ref{tivix}). Integrate the latter over a thin 3--dimensional slice of width ${\rm d}n$ bounded by two equipotentials ${\cal S}_1$ and ${\cal S}_2$. Now the Planck length $L_P$ is extremely small, so we can safely set ${\rm d}n=L_P$, while the two screens ${\cal S}_1$ and ${\cal S}_2$ will not differ appreciably in their surface area. Then the volume integral of the left--hand side of (\ref{tivix}) very approximately equals $2\pi k_BL_P\int_{\cal S}{\rm d}^2a\,\hat T$. On the right--hand side, let us first integrate ${\rm d}\hat U/{\rm d}n$ along the normal direction, to obtain $L_P\hat U({\cal S}_2)-L_P\hat U({\cal S}_1)$. We can take the origin for the potential function such that it will vanish on ${\cal S}_1$. The remaining term is the surface integral $L_P\int_{\cal S}{\rm d}^2a\,\hat U({\cal S})$. The integrand can be pulled past the integration sign because ${\cal S}$  is an equipotential surface, thus yielding $L_P\hat U({\cal S})\int_{\cal S}{\rm d}^2a$. This latter integral equals the surface area $A({\cal S})$ of the screen, and (\ref{deficuanto}) follows as claimed.

Taking the expectation value, in the entropic eigenstates $\phi_-$, of the operator equation (\ref{deficuanto}), produces the thermodynamical expression for the equipartition theorem:
\begin{equation}
\frac{k_B}{2}\int_{\cal S}{\rm d}^2a\, \langle\hat T\rangle=\frac{A({\cal S})}{4\pi}\langle\hat U ({\cal S})\rangle.
\label{equinova}
\end{equation}

\subsection{Planck {\it vs}\/. Boltzmann, or $\hbar$ {\it vs}\/. $k_B$}\label{pvsbltzmnn}

Planck's quantum of action $\hbar$ gets replaced, in the entropic picture, with Boltzmann's constant $k_B$. This explains the presence of $\hbar$ in the entropic derivation of classical gravity (Newton's and Einstein's) given in ref. \cite{VERLINDE}: by the correspondence (\ref{korres}), {\it the presence of $\hbar$ is an unavoidable consequence of the presence of $k_B$, and viceversa}\/. We find  this dichotomy between the energy and the entropy representations very suggestive---it appears to be a sort of complementarity principle, in Bohr's sense of the word. For example, this dichotomy allows one to write a quantum of energy in the form $E=\hbar\omega$, or else in the alternative form $E=Ck_BT$ ($C$ being a dimensionless number). It also allows one to express a quantum of entropy in the form $S=\hbar\omega/T$, or else as $S=2\pi k_B$. This dichotomy exchanges frequency $\omega$ with temperature $T$, thus time $t$ maps to inverse temperature $T^{-1}$, which is reminiscent of the Tolman--Ehrenfest relation \cite{TOLMAN} and also of thermal time \cite{ROVELLI}.

\subsection{The second law of thermodynamics, revisited}\label{tflwotrev}

As a minor technical point, we have restricted our analysis to closed holographic screens enclosing a finite 3--dimensional volume. Quantum--mechanically this corresponds to normalisable states in the energy representation. Nonnormalisable states correspond to open holographic screens without a boundary (thus having an infinite surface area and enclosing an infinite volume). Our analysis can be extended to the latter by replacing absolute quantities with densities (per unit surface or unit volume as the case may be). The connection with the second law of thermodynamics comes about as follows. The second law of thermodynamics, $\Delta S\geq 0$, lies hidden within the quantum theory. Of course, one can derive it from statistical mechanics, but our purpose here is the opposite. We have seen that the domain of the reduced entropy $s$ is the half axis $s\geq 0$, and that this fact led to the entropic Hilbert space $L^2[0,\infty)$ (instead of $L^2(\mathbb{R})$) for the wavefunctions $\psi_-(s)$. All this is a quantum--mechanical rewriting of the second law. One could ask, under what conditions will the entropic coordinate $\rho$ be nonnegative? This is certainly the case when the holographic screens are all closed, but what happens in case they are open? The geometry of the screens is dictated by the potential $U$. If the latter has flat directions, then its equipotentials will no longer be closed surfaces---instead they will have an infinite surface area and will enclose an infinite volume. As mentioned above, one appropriately replaces quantities like entropy and energy with the corresponding densities. However, the corresponding screens must be such that the normal coordinate to their bright side, $\rho$, runs over the half axis $\rho\geq 0$. This latter condition will be satisfied whenever the potential is such that it possesses a centre of force, or an axis, or a plane, or possibly a more general surface of symmetry, with respect to which one can define a nonnegative normal coordinate. This appears to be the case in all physically interesting situations, thus staying in agreement with the second law of thermodynamics. Only the free particle lacks a {\it canonical}\/ definition of a normal coordinate---but then again the second principle holds in the form $\Delta S=0$, due to the absence of forces.

\section{Discussion}\label{disco}

\subsection{Quantum mechanics as a holographic, emergent phenomenon}\label{qmahemephe}

Classical thermodynamics can be conveniently expressed in either of two equivalent languages, respectively called the energy representation and the entropy representation \cite{CALLAN}. Here we have argued that quantum mechanics as we know it ({\it i.e.}, on spacetime) corresponds to the energy representation, while quantum mechanics beyond a holographic screen ({\it i.e.}, in the absence of spacetime) corresponds to the entropy representation. In this paper we have developed the formalism of entropic quantum mechanics and placed it in correspondence with that of standard quantum mechanics on spacetime.

In particular, we have formulated the entropic uncertainty principle (\ref{ungewiss}) for the (reduced) entropy variable $s$ that the entropic wavefunction $\psi_-(s)$ (sometimes also denoted $R(\rho)$) depends on; see (\ref{hijosdeputa}). The latter arises as the result of factoring out the part of the wavefunction that depends on the tangential coordinates to the screen, the normal coordinate being proportional to the entropy itself. We have also written down a differential equation satisfied by the entropic wavefunction, that one may well call the entropic Schroedinger equation; see (\ref{raddius}).

Moreover, we have identified the explicit expression (\ref{gemuetlich}) as corresponding to the entropic wavefunction in the semiclassical limit $k_B\to 0$. There is a nice map, given by (\ref{korres}), between the semiclassical wavefunction in the energy representation and the corresponding semiclassical wavefunction in the entropy representation. This map exchanges the classical action integral with the entropy of the screen, while at the same time introducing a relative factor of $i$. It also exchanges Planck's constant $\hbar$ with Boltzmann's constant $k_B$. In so doing, this map succeeds in explaining why Planck's constant $\hbar$ had to appear in the derivation of classical gravity (Newton's and Einstein's) given in ref. \cite{VERLINDE}. Namely, the presence of $\hbar$ is an inescapable consequence of the presence of $k_B$, and viceversa, since $\hbar$ is required by the energy representation, while $k_B$ is required by the entropy representation.

If spacetime is an emergent phenomenon, then everything built on it necessarily becomes emergent \cite{FREUND}. This applies to quantum mechanics in particular. However, in the entropy representation developed here, the emergence property of quantum mechanics becomes a much sharper feature. Indeed, one usually associates entropy with lack of information, while energy ({\it e.g.}, a sharp energy eigenvalue) is thought of as providing definite information. Now the correspondence (\ref{korres}) implies that, if the entropy representation is emergent, then so is the energy representation, and viceversa.  In this sense, the information content carried by entropy is no more diffuse than that carried by energy, nor is the information encoded by energy more sharply defined than that encoded in entropy. In other words, the correspondence (\ref{korres}) confirms what we already knew from other sources---namely, that quantum mechanics is definitely an emergent phenomenon.

We have also succeeded in writing a holographic dictionary between quantum mechanics, on the one hand, and thermodynamics, on the other. An analogous holographic dictionary was presented, in the gravitational case, in ref. \cite{VERLINDE}. Some key entries in this gravitational/thermodynamical dictionary are summarised in eqns.  (\ref{sabinebiene}), (\ref{tivi}) and (\ref{defitermonova}), preceded by the equivalence between statements {\it i)} and {\it ii)} of section \ref{aholdic}. As a novelty, here we have presented the corresponding entries in our quantum--mechanical/thermodynamical dictionary. These entries include the equivalence between the analogues of statements {\it i)} and {\it ii)} of section \ref{aholdic}. In our setup, this is expressed in the assertion that the energy representation of quantum mechanics (statement {\it i)}) is equivalent to the entropy representation of quantum mechanics (statement {\it ii)}). Further entries in this dictionary of equivalences are the analogues of eqns. (\ref{sabinebiene}), (\ref{tivi}) and (\ref{defitermonova}), respectively given by our eqns. (\ref{lieber}), (\ref{tivix}) and (\ref{deficuanto}) when working in the energy representation. Our eqns. (\ref{lieber}), (\ref{tivix}) allow one to define an entropy field and a temperature field as (operator--valued) functions on $\mathbb{R}^3$, whereas (\ref{deficuanto}) is a reexpression of the first law of thermodynamics and of the equipartition theorem. Their respective vacuum expectation values give rise to the corresponding equations in the entropy representation, (\ref{bhache}), (\ref{estado}) and (\ref{equinova}), where the space dependence  disappears. Their respective interpretations are  the proportionality between the area and the entropy of the screen (the Bekenstein--Hawking law), the thermodynamical equation of state of the screen (the Unruh law), and the equipartition theorem.

\subsection{Quantum mechanics in the absence of spacetime}\label{qmitabsospct}

Entropic quantum mechanics can be thought of as describing quantum mechanics {\it in the absence of spacetime}\/. This latter statement must be understood as meaning that the tangential coordinates to the holographic screens, as well as functions thereof, have been factored out, while the normal coordinate and functions thereof remain---though no longer as a {\it spatial coordinate}\/, but rather as a {\it measure of entropy}\/. This viewpoint is motivated in eqn. (\ref{verlindeq}), that we have borrowed directly from \cite{VERLINDE}. Now in the absence of time there is no Hamiltonian. In the absence of space there are also no paths to sum over {\it \`a la}\/ Feynman. One might thus conclude that there can be no quantum mechanics in the absence of spacetime. This is however not true, as shown here and as shown also by independent analyses. For example, quantum mechanics without spacetime has been proposed as a case for noncommutative geometry \cite{SINGH1, SINGH2, NOI2}. Without resorting to noncommutative geometry, one can also argue as follows.

We have seen that the Hilbert space of entropic quantum states is $L^2[0,\rho_j)$ for an observer who perceives space terminating at the screen ${\cal S}_j$, and $L^2[0,\infty)$ for an  observer who perceives no screen at all, or horizon. Given the two screens ${\cal S}_j$ and ${\cal S}_k$, respectively located at $\rho=\rho_j$ and $\rho=\rho_k$ with $\rho_j<\rho_k$, it holds that the two spaces $L^2[0,\rho_j)$ and $L^2[0,\rho_k)$ are unitarily isomorphic because both are infinite--dimensional and separable \cite{THIRRING}. Now let $\rho_k\to\infty$. The isomorphism between $L^2[0,\rho_j)$ and $L^2[0,\infty)$, plus the identification (\ref{hijosdeputa}) between entropy and normal coordinate, allows the observer who perceives the screen ${\cal S}_j$ to extend his wavefunctions $R(\rho)$ beyond his boundary at $\rho_j$. His wavefunctions are now understood as $\psi_-(s)$, {\it i.e.}, as functions of the reduced entropy $s$---indeed the latter is not bounded from above. It is in this sense that this second observer can be said to be doing {\it quantum mechanics in the absence of spacetime}\/.

It is right to observe that the unitary isomorphism between the two different realisations of the entropic Hilbert space, $L^2[0,\infty)$ and $L^2[0,\rho_j)$, need not map the semiclassical regime of the one into the semiclassical regime of the other, nor the strong--quantum regime of the one into the corresponding regime of the other. An analogous statement applies to the spaces $L^2[0,\rho_j)$ and $L^2[0,\rho_k)$ corresponding to the screens ${\cal S}_j$, ${\cal S}_k$. The observation just made will become relevant in section \ref{opque}.

\subsection{Open questions}\label{opque}

We can summarise our conclusions so far with the assertion that entropic quantum mechanics is a holographic phenomenon, as emergent as spacetime itself. To round up our discussion we would like to present some thoughts of a more speculative nature.

As a first thought we would like to state that {\it entropic quantum mechanics is an observer--dependent phenomenon}\/. That measurement disturbs any quantum system is, of course, a basic tenet of quantum mechanics. The statement just made, however, refers to something different.  The concept that quantum mechanics is observer--dependent has also appeared, in different guises, in  \cite{VAFA, SINGH3, SINGH4} under the name of {\it duality}\/. Under duality one understands that {\it the notion of classical vs. quantum is relative to which theory one measures from}\/ (see section 6 of  ref. \cite{VAFA}). This is also the interpretation advocated in refs. \cite{NOI4} by one of the present authors.

An idea that lies close to the above notions is the statement that {\it the entropy of a horizon is an observer--dependent quantity}\/ (see section 3 of ref. \cite{PADDY1}). In view of our correspondence (\ref{korres}), this latter assertion turns out to be equivalent to the one above defining duality.

Thus the statement that quantum mechanics is observer--dependent,  is an equivalent reexpression of duality, {\it i.e.}, of the relativity of the notion of a quantum. In the entropic picture developed here, this relativity presents itself as the different realisations of the entropic Hilbert space, explained in section \ref{qmitabsospct}. Equivalently, this relativity of the notion of a quantum arises here as the relativity of the entropy.

The previous statements may at first sound surprising. Classic treatises such as, {\it e.g.}, ref. \cite{TOLMAN}, teach that the Lorentz transformation laws for the heat energy and the temperature are such that their ratio (the entropy) is a scalar. Moreover, in principle one expects physical constants such as $k_B$  and $\hbar$ to be observer--independent. However, let us note that a totally analogous phenomenon has been reported in refs. \cite{PADDY00, PADDY0, PADDY1, PADDY2, PADDY3}, where the entropy of the screen has been argued to be an observer--dependent quantity. That the entropy of a thermodynamical system becomes an observer--dependent quantity has also been concluded in an information--theoretical context \cite{NENTROPY}. Upon transforming back to the energy representation, the dependence just described can be recast as the dependence of Planck's constant $\hbar$ upon the observer. Exactly this latter conclusion concerning $\hbar$ has been reported in \cite{VOLOVIK}.

Given that the equations of motion for Einstein's gravity can be recast as thermodynamical equations of state, it has been claimed  that the canonical quantisation of gravity makes as little sense as {\it quantising sound waves in air}\/ \cite{JACOBSON}. This remark makes it clear that quantising Einstein's gravity may be attempting to quantise the wrong classical theory, but it casts no doubt yet on the validity of quantum theory. However, doubts concerning the microscopic fundamentality of the latter arise once one realises that {\it quantum theory, too, is a thermodynamics in disguise}\/...

\noindent
{\bf Acknowledgements}  J.M.I. thanks Max--Planck--Institut f\"ur Gravitationsphysik,\\ Albert--Einstein--Institut (Golm, Germany), for hospitality extended a number of times over the years. This work has been supported by Universidad Polit\'ecnica de Valencia under grant PAID-06-09.\\
{\it Dich st\"ore nichts, wie es auch weiter klinge,\\ schon l\"angst gewohnt der wunderbarsten Dinge.\\---Goethe.}

\end{document}